\documentclass[12pt]{amsart} 
\usepackage[mathscr]{eucal}
\usepackage{amsmath,amsfonts}

\parskip=\smallskipamount

\newtheorem{theorem}{Theorem}
\newtheorem{definition}[theorem]{Definition}
\newtheorem{proposition}[theorem]{Proposition}

\newtheorem{remark}[theorem]{Remark}

\makeatletter
\@addtoreset{equation}{section}
\makeatother

\newcommand{\cA}{{\mathcal A}}

\newcommand{\cD}{{\mathcal D}}

\newcommand{\cH}{{\mathcal H}}

\newcommand{\cL}{{\mathcal L}}
\newcommand{\cM}{{\mathcal M}}

\newdimen\expt
\def\boxit#1{\setbox0\hbox{$\displaystyle{#1}$}
      \hbox{\lower.4\expt
 \hbox{\lower3\expt\hbox{\lower\dp0
      \hbox{\vbox{\hrule height.4\expt
 \hbox{\vrule width.4\expt\hskip3\expt
      \vbox{\vskip3\expt\box0\vskip2\expt}%
 \hskip3\expt\vrule width.4\expt}\hrule height.4\expt}}}}}}

\begin{document}
\pagestyle{plain}

\bigskip

\title 
{Parametrizations of Positive Matrices With Applications}
\author{M. C. Tseng} \author{H. Zhou} \author{V. Ramakrishna}

\address{Department of Mathematics \\
  University of Texas at Dallas \\
  Box 830688, Richardson, TX 75083-0688, U. S. A.}
\email{\tt vish@utdallas.edu}

\begin{abstract}
The purpose of this work is twofold. The first is to survey some
parametrizations of positive matrices which have found applications
in quantum information theory. The second is to provide some more
applications of a parametrization of quantum states and channels
introduced by T. Constantinescu and the last author, and thereby to
provide further evidence of the utility of this parametrization.
This work is dedicated to the memory of our colleague and teacher,
the late Professor T. Constantinescu.
\end{abstract}

\maketitle

\section{Introduction}

Positive matrices play a vital role in quantum mechanics and its
applications (in particular, quantum information processing).
Indeed the two basic ingredients in the theory of quantum
information, viz., quantum states and quantum channels 
involve positive matrices. See, for instance, \cite{nie,preskill}. 
Thus, a study of parametrizations of positive matrices seems 
very much warranted.
In particular, the very useful
Bloch sphere picture, \cite{nie,preskill}, for the quantum state of a qubit has
prompted several attempts at the extension of this picture to
higher dimensions. In the process, several groups of researchers have
looked into the question of finding tractable characterizations of
positive matrices, which could lead to useful parametrizations of
positive matrices, \cite{bird,Ki,leahy,Za,tivin}.

This paper is organized as follows.
In the next section we set up basic notation and also point
out some sources for positive matrices in quantum mechanics and its
applications. The third section introduces six (perhaps well-known)
characterizations of
positive matrices, and reviews some putative parametrizations of states of
qudits. In the 
next section, we review a parametrization proposed in \cite{tivin},
reiterating its utility. The final section offers two more applications
of the parametrization in \cite{tivin}. The first concerns Toeplitz states,
i.e., density matrices which are also Toeplitz. The second investigates
constraints imposed on relaxation rates of an open quantum system by
the requirement of complete positivity.

\section{Sources of Positive Matrices in Quantum Theory}
Let us recall that a matrix is positive semidefinite (positive, for short)
if $z^{*}Pz\geq 0$ for all $z\in C^{n}$. One can easily extend this
definition to infinite positive matrices. In effect such a matrix
is what is called a positive kernel, \cite{Co}, viz., a map
$K: N_{0}\times N_{0} \rightarrow C$, where $N_{0}$ is the set of non-negative
integers, with the property that for each $n > 0$, and each 
choice $p_{1}, \ldots , p_{n}$ in $N_{0}$ and each choice 
$z_{1}, \ldots , z_{n}$ of elements of $C$ we have
\[
\sum_{i,j=1}^{n}K(p_{i}, p_{j})\bar{z_{i}}z_{j} \geq 0
\]

Positive matrices intervene in at least two of the basic ingredients
of quantum mechanics and quantum information theory viz., quantum
states and quantum channels. There are, of course, more sources for
positive matrices, but, for reasons of brevity, we will confine
ourselves to discussing states and channels. 

The state of a $d$-dimensional quantum system is described by
a $d\times d$ positive density matrix of trace 1, that is, a
positive element of trace $1$ in the algebra $\cM _d$ of complex $d\times d$
matrices. States described by rank one density matrices are called pure
states.

A quantum channel is a completely positive map
$\Phi :\cA \rightarrow \cL({\cH})$
from a $C^*$-algebra $\cA $ into the set $\cL(\cH)$ of all bounded
linear operators on the Hilbert space $\cH$ (in the situations most
frequently met in quantum information processing, $\cA = \cM _d$
and $\cL(\cH) = \cM _{d^{'}}$). 
By the Stinespring theorem, \cite{Pa}, Theorem~4.1, such a map
is the compression of a $*$-homomorphism. For $\cA =\cM _d$,
there is a somewhat more explicit
representation, given in \cite{Ch} (see also \cite{Jam}).
Thus, $\Phi :\cM _d\rightarrow
\cL(\cH)$ is completely positive if and only if
the matrix
\begin{equation}\label{sphi}
 S=S_{\Phi }=\left[\Phi (E_{k,j})\right]_{k,j=1}^d
 \end{equation}
 is positive, where $E_{k,j}$, $k,j=1,\ldots ,d,$
 are the standard matrix units of $\cM_d$. Each
 $E_{k,j}$ is a $d\times d$ matrix
consisting
of $1$ in the $(k,j)th$ entry and zeros elsewhere.

\begin{remark}
{\rm Usually one requires a quantum channel to satisfy two additional
requirements: i) $\Phi$ be trace preserving, and/or ii) $\Phi$ be unital. 
}
\end{remark}

A Kraus operator representation of a completely positive map
is a (non-unique) choice of operators $V_{i}$ such that one
can express the effect of $\Phi$ via
\[
\Phi (\rho ) = \sum_{i=1}^{r}V_{i}\rho V_{i}^{*}
\]
 
Usually only the non-zero $V_{i}$ are taken into account in the above equation
(though sometimes it is
convenient to ignore this convention). 

Then $\Phi$ is trace-preserving iff $\sum_{i=1}^{r}V_{i}^{*}V_{i} = {\mbox Id}$,
while $\Phi$ is unital iff 
$\sum_{i=1}^{r}V_{i}V_{i}^{*} = {\mbox Id}$. These properties can also
be verified (without any reference to Kraus representations) by computing
the partial traces  of $S_{\Phi}$
viewed as an unnormalized state (see \cite{vv}).
  
All choices of Kraus operator representations for $\Phi$ come from
square-roots of $S_{\Phi}$, i.e., matrices $T$ such that
$S_{\Phi} = TT^{*}$. One then obtains the $V_{i}$ from the $i$th column
of $T$ by reversing the {\it vec} operation, \cite{tim,vv}. 
Recall that the {\it vec}
operator associates to a $d\times e$ matrix, $V$, a vector in $C^{de}$ obtained
by stacking the columns of $V$. It is precisely because of lack of
uniqueness in the square roots of $S_{\Phi}$ that the Kraus operator
representation of $\Phi$ is non-unique.
 
We should point out that some of the definitions for quantum channel
notions used in \cite{tivin}, though equivalent to the standard ones
(i.e., the ones used here), are different.    
                                                                                
\section{Characterizations of Positive Matrices}
All positive matrices are Hermitian (unlike the real case, the definition
of a positive matrix automatically forces Hermiticity). There are several
characterizations of positive matrices as a subclass of Hermitian matrices.
Some of these yield useful parametrizations of positive matrices.

The following theorem, which for the most part is standard textbook material
(see, for instance, the classic \cite{hornj}),
reviews some of these characterizations. 

\begin{theorem}
Let $P$ be a Hermitian matrix. Then the following are equivalent:
\begin{itemize}
\item {\bf P1} $P$ is positive.
\item {\bf P2} All the eigenvalues of $P$ are non-negative.
\item {\bf P3} There is an upper-triangular matrix $T$ such that
$P = T^{*}T$ (Cholesky decomposition)
\item {\bf P4} All principal minors of $P$ are non-negative.
\item {\bf P5} Let $p(t) = t^{n} + \sum_{i=0}^{n-1} (-1)^{i}b_{i}t^{n-i}$
be the characteristic polynomial of $P$. Then $b_{i}\geq 0$, for all $i$.
\item {\bf P6} There is a Hermitian matrix $H$ such that $P = H^{2}$.
\end{itemize}
\end{theorem}

\begin{remark}
i) {\bf P3} {\rm is normally mentioned only for positive definite matrices 
in the bulk of the literature. However, a limiting argument shows that it
is valid for positive semidefinite matrices as well.} ii) {\rm That}
{\bf P5} {\rm is equivalent to} {\bf P2} {\rm is just a consequence
of Descartes' rule of signs.} iii) {\bf P4} {\rm should be folklore.
Quite surprisingly, we were
unable to find any source where} {\bf P4} {\rm is stated explicitly  
(even in a venerable text such as \cite{hornj}). 
Since a similar statement for  
positive definite matrices (viz., positive
definiteness is equivalent to the positivity of the {\it leading} principal
minors) is well documented and we have seen this statement occassionally 
incorrectly applied to positive semidefinite matrices, 
we will include a brief
proof here. Clearly if $P$ is positive, all principal submatrices
of $P$ are positive, and hence all principal minors are non-negative.
Conversely suppose all principal minors of $P$ are non-negative.
Since the coefficients $b_{i}$ of the characteristic polynomial of
any matrix are just the sum of all the $i\times i$ principal minors
of $P$, it follows that $b_{i}\geq 0$. Hence $P$ is positive.}
\end{remark}  

Whilst the above conditions are equivalent to positivity, they typically
do not lead to useful parametrizations of positive matrices. For instance,
parametrizing $P$ by its eigenvalues only describes the $U(n)$ orbit
to which $P$ belongs. For the same reason one cannot parametrize
$P$ by the coefficients $b_{i}$ of the characteristic polynomial
$p(t)$.

However, one can turn these characterizations into {\it potential}
parametrizations. To illustrate this consider the problem of
parametrizing quantum states in dimension $d$, i.e., $d\times d$
positive matrices with unit trace. The standard starting point is
to represent a state $\rho$ via
\begin{equation}
\label{innd}
\rho = \frac{1}{d}(I_{d} + \sum_{i=1}^{d^{2}-1}\beta_{i}\lambda_{i})
\end{equation}
Here $\beta_{i}\in R$ and the $\lambda_{i}$
form an orthogonal basis
for the space of traceless Hermitian matrices, specifically
${\mbox Tr} (\lambda_{i}\lambda_{j}) = 2\delta_{ij}$.
One typical choice is the
so-called generalized Gell-Mann matrices, \cite{Ki,Za}.
This basis is obtained from the matrices $E_{kj}, k,j=1, \ldots , d$
($E_{kj} = e_{k}e_{j}^{*}$) via the following construction:
$$f^d_{k,j}=E_{k,j}+E_{j,k},\quad k<j,$$
$$f^d_{k,j}=\frac{1}{i}\left(E_{j,k}-E_{k,j}\right),\quad k>j,$$
$$h_1^d=I_d,\quad h_k^d=h_k^{d-1}\oplus 0, \quad 1<k<d, \quad
h_d^d=\sqrt{\frac{2}{d(d-1)}}\left(h_1^{d-1}\oplus (1-d)\right).$$
These matrices, $f^d_{k,j}, h_1^d,  h_k^d, h_d^d$ together form one
choice of the $ \{\lambda_{i}, I_{d}\}$ basis for the
space of $d\times d$ Hermitian matrices. When $d=2$ this is precisely
the Pauli matrix basis. When $d=3$ one gets the usual Gell-Mann matrices.

With Equation (\ref{innd}) as the starting point one can restrict the
vector $\beta = (\beta_{1}, \ldots , b_{d^{2}-1}) \in R^{d^{2}-1}$ 
to satisfy any of the 
characterizations {\bf P1 - P6}. In principle, this provides
a bijection from a subset of $R^{d^{2}-1}$, say  $D_{\beta}$, 
to the space of $d\times d$
density matrices. This is precisely what is proposed simultaneously
in \cite{Ki,bird} for the characterization {\bf P5}. However, now by
conservation of difficulty, the burden of the analysis of quantum
states in dimension $d> 2$ is shifted to obtaining a concrete analysis
of the subset $D_{\beta}$. In particular, these do not lead to
easily computed parametrization of quantum states (cf., the conclusions
section of \cite{Ki}). Interestingly enough each of these characterizations
leads precisely to the Bloch sphere picture when $d=2$, as we encourage
the reader to verify. However, this approach has some utility in higher
dimensions as well.
For instance, depending on which characterization one uses, it 
is at least possible to be more precise about the set of pure states
(i.e, rank one states). We shall explain this via the characterization
{\bf P6} because pure states are precisely those states, $\rho$ for
which $\rho^{2} = \rho$, and this fits in nicely with {\bf P6}.

In order to state a precise result, let us introduce the tensor
$d_{kli}$ obtained from considering the Jordan structure of the
$\lambda_{i}$. Specifically, if $\{\lambda_{k}, \lambda_{l}\}$
denotes the Jordan commutator of $\lambda_{k}, \lambda_{l}$, then
\[
\{\lambda_{k}, \lambda_{l}\} = \lambda_{k}\lambda_{l} + \lambda_{l}\lambda_{k}
=  \frac{4}{d}I_{d}\delta_{kl} + \sum_{i=1}^{d^{2}-1}d_{kli}\lambda_{i}
\]
We use the $d_{kli}$ to introduce an operation amongst vectors $x,y\in
R^{d^{2}-1}$, via
\[
x\cup y =
(\sum_{j,k = 1}^{d^{2} - 1} d_{1jk}x_{j}y_{k}, 
\sum_{j,k = 1}^{d^{2} - 1} d_{2jk}x_{j}y_{k}, \ldots ,
\sum_{j,k = 1}^{n^{2} - 1} d_{ijk}x_{j}y_{k}, \ldots ) 
\]
$x\cup y$ is thus a vector in $R^{d^{2}-1}$. We can now state
\begin{proposition}
{\rm Every density matrix can be represented in the form in Equation
(\ref{innd}) with
$\beta = \frac{2\kappa}{d}\beta_{0} + \frac{\beta_{0}\cup\beta_{0}}
{d}$, where $\beta_{0}$ is any vector in $R^{d^{2}-1}$ with
$\mid\mid\beta_{0}\mid\mid^{2}\leq \frac{d^{2}}{2}$ and $\kappa =
+ \sqrt{\frac{d^{2}- 2 \mid\mid\beta_{0}\mid\mid^{2}}{d}}$.
Conversely any Hermitian matrix admitting such a representation is
necessarily a density matrix. $\rho$ is pure precisely
if it can be represented in the form in Equation \ref{innd})
with $<\beta , \beta > = \frac{d^{2}-d}{2}$ and $(d-2)\beta =
\beta \cup \beta$.}
\end{proposition}

The proof is straightforward. Since $\rho = H^{2}$ and $H$ itself can
be expanded as a linear combination of $I_{d}$ and the $\lambda_{i}$
(albeit with the coefficient of $I_{d}$ different from $\frac{1}{d}$),
the first part of the result follows from the linear independence of
$\{I_{d}, \lambda_{i}\}$. For the second part, we represent $\rho$ as in
Equation (\ref{innd}) and equate it to its square. 

Once again, the difficulty is in the analysis of states which are not
pure.  It is worth mentioning that
the pure state condition in the above proposition is essentially 
the same as that obtained
from the characterization {\bf P5} (for a pure state the characteristic
polynomial is $p(t) = t^{d} + (-1)^{d}t^{d-1}$, i.e., $b_{1} = 1,
b_{i} =0, i\geq 2$).

It should be pointed out that even an analysis of the pure state conditions
is far from trivial. The condition $(d-2)\beta =
\beta \cup \beta$ is vacuously true when $d=2$ (since the Pauli matrices
anti-commute). For $d\geq 3$, this condition imposes genuine restrictions.
It is an interesting problem to find an orthogonal basis for 
the space of Hermitian
matrices (the generalized Gell-Mann matrices form just one amongst many)
which is close to ``abelian", i.e, one for which many of the $d_{kli}$
vanish, to facilitate the analysis of the condition $(d-2)\beta =
\beta \cup \beta$. 

In contrast, the parametrization discussed in the next section
yields a very simple characterization of pure states.
  
\section{A different parametrization of positive matrices}
In this section we recall informally the main result of \cite{tivin} on the
parametrization of positive matrices. In order to do that a few preliminary
definitions and notions are needed.
                                                                                
\noindent To any contraction $T$, one defines its defect operator via
\[
D_T=(I-T^*T)^{1/2}
\]
Here $M^{*}$ is the adjoint of an operator (when $M$ is a scalar, this
is merely complex conjugation).

To such a contraction one can also associate a certain unitary operator,
called the {\it Julia operator} of $T$ via
\begin{equation}\label{rotele}
 U(T)=\left[
 \begin{array}{cc}
 T & D_{T^*} \\
 D_T & -T^*
 \end{array}
 \right].
 \end{equation} 
Thus, $U(T)$ is a unitary dilation of $T$.

If we are given a family of contractions $\Gamma_{kj}, j\geq k$ with
$\Gamma_{kk} = 0$ for all $k$, then we associate to it a family of
unitary operations via the Julia operator construction as follows.
We first let $U_{k,k}= {\mbox Id}$, while for $j > k$ we set
$$U_{k,j}=U_{j-k}(\Gamma _{k,k+1})U_{j-k}(\Gamma _{k,k+2})\ldots
 U_{j-k}(\Gamma _{k,j})(U_{k+1,j}\oplus I_{\cD _{\Gamma ^*_{k,j}}}),
$$
where $$
U_{j-k}(\Gamma _{k,k+l})=I\oplus U(\Gamma _{k,k+l})\oplus I.
$$

To a family of contractions, $\Gamma_{k, j}$ one can associate
a row contraction via
\[
R_{k,j}=\left[
 \begin{array}{cccc}
 \Gamma _{k,k+1}, & D_{\Gamma ^*_{k,k+1}}\Gamma _{k,k+2},
 & \ldots ,& D_{\Gamma ^*_{k,k+1}}\ldots D_{\Gamma ^*_{k,j-1}}\Gamma _{k,j}
 \end{array}\right]
\]
and a column contraction via
\[
C_{k,j}=\left[
 \begin{array}{cccc}
 \Gamma _{j-1,j}, & \Gamma _{j-2,j}D_{\Gamma _{j-1,j}},
 & \ldots ,& \Gamma _{k,j}D_{\Gamma _{k+1,j}}\ldots D_{\Gamma _{j-1,j}}
 \end{array}\right]^t,
\]
where $"t"$ stands for matrix transpose.
For more details on the ranges and domains of these operators see \cite{tivin}.

Then the main theorem regarding positive matrices can be stated
informally as follows (for a precise statement, especially concerning
the ranges and domains of all operators involved, see \cite{tivin})

\begin{theorem}
\label{csp}
{\rm
The matrix $S=\left[S_{k,j}\right]_{k,j=1}^d$
as above, satisfying $S_{jk}^{*} = S_{kj}$, is positive if and only if
i) $S_{kk} \geq 0, k=1, \ldots, d$ and ii) there exists a family
$\{\Gamma _{k,j}\mid k,j=1,\ldots ,d, k\leq j\}$ of contractions
such that
$\Gamma _{k,k}=0$ for $k=1,\ldots ,d$ 
valid, and
\begin{equation}\label{relation}
 S_{k,j}=L^*_{k,k}(R_{k,j-1}U_{k+1,j-1}C_{k+1,j}+
 D_{\Gamma ^*_{k,k+1}}\ldots
 D_{\Gamma ^*_{k,j-1}}\Gamma _{k,j}
 D_{\Gamma _{k+1,j}}\ldots
 D_{\Gamma _{j-1,j}})L_{j,j}.
 \end{equation}
where $L_{k,k}$ is any square root of $S_{kk}$. 
}
\end{theorem}

\begin{definition}
{\rm The contractions $\Gamma_{k,j}$, with $j > k$, will be called}
the Schur-Constantinescu
{\rm parameters of $S$}.
\end{definition}
                                                                                
These parameters were first discovered for Toeplitz matrices by
Schur, \cite{ISchur}, albeit in the guise of a problem about
power series which are bounded in the unit circle. In our humble opinion,
it was our late colleague and teacher, T. Constantinescu, who championed
the study of these parameters to cover all positive matrices
(more generally to 
matrices with displacement structure, \cite{CSK}) and most adroitly brought to
fore many of their interesting features. Therefore,
we have chosen to call these parameters, the Schur-Constantinescu parameters,
in his honour.

We will illustrate Theorem (\ref{csp}) via the case of $3\times 3$
positive matrices.
                                                                                
Thus, let $S=\left[\begin{array}{ccc}
S_{11} & S_{12} & S_{13} \\
S^*_{12} & S_{22} & S_{23} \\
S^*_{13} & S^*_{23} & S_{33}
\end{array}\right]$
be a positive matrix. Then $S_{ii} > 0$ and let us pick
$L_{ii}$ as the positive square roots of $S_{ii}$. In this case
$L_{ii}^{*} = L_{ii}$. Then per Theorem
~\ref{csp}, there are complex numbers $\Gamma_{12}, \Gamma_{13},
\Gamma_{23}$ in the unit disc such that
gives:
$$
S_{12}=L^*_{11}\Gamma _{12}L_{22},
$$
$$
S_{23}=L^*_{22}\Gamma _{23}L_{33},
$$
$$
S_{13}=L^*_{11}\left(\Gamma _{12}\Gamma _{23}+
D_{\Gamma ^*_{12}}\Gamma _{13}
D_{\Gamma _{23}}\right)L_{33},
$$
                                                                                
Note that there is a recursive procedure to determine the $\Gamma_{kj}$.
The first and the second equations yield $\Gamma_{12}, \Gamma_{23}$
from quantities already known, while the last equation yields
$\Gamma _{13}$ from quantities already determined at the first two equations.
                                                                               
Whilst, the Schur-Constantinescu paramters are defined directly in terms
of the entries of $S$, one could also seek expressions for them in terms
of the vector $\beta$ of Equation (\ref{innd}) (i.e, when $S$ is a density
matrix). See \cite{CR,tivin} for such expressions. In particular, for
$d=2$ the analogue of the Bloch sphere is now a cylinder.

It is appropriate to make several comments about these parameters at this
point:
\begin{itemize}
\item {\bf C1} As can be expected from the form
of Equation (\ref{relation}),
Theorem (\ref{csp}) is valid for operator matrices, i.e.,
matrices whose entries are matrices or even operators in infinite-dimensional
spaces, i.e., for elements of $\cM _d \otimes \cL (\cH )$,
with $\cH$ allowed to be infinite-dimensional.
In fact, one can easily extend the result to infinite matrices
with  (possibly infinite-dimensional) operator entries.
                                                                                
\item {\bf C2} 
Though we only called the $\Gamma_{kj}$ as the Schur-Constantinescu
parameters, a full parametrization is provided by the
$\frac{d(d-1)}{2}$ contractions $\Gamma_{kj}, k < j$ and the
$L_{ii}, i=1, \ldots , d$. In the case of scalar valued matrices,
i.e., when $\cH = C$, we thus get the right count of $d^{2}$ real
parameters.
Note the $\Gamma_{kk} = 0$ are just some fake parameters,
included in the statement of the theorem to avoid an artificial separation
of the $j = k+1$ case from that for other values of $j$.
                                                                                
\item {\bf C3} Since the $L_{ii}$ are allowed to be any choice of square root
of $S_{ii}$ (i.e., $S_{ii} = L_{ii}L_{ii}^{*}$), the parametrization
will be different for different choices of the $L_{ii}$. A most
natural choice would be the Cholesky factorization of $S_{ii}$.
In fact, as described in \cite{tivin}, there is an algorithmic proof
of Theorem (\ref{csp}) which automatically yields the Cholesky factorization
of $S$. In the infinite-dimensional case, some of the algorithmic
flavour of the proof is lost.
                                                                                
\item {\bf C4} While, Equation (\ref{relation}) in Theorem (\ref{csp}) is
nonlinear and looks quite complicated, there is an iterative feature
to it (as mentioned in the $3\times 3$ example given before), inasumch
as in each equation there is just one of the $\Gamma_{kj}$ being solved for.
It is precisely because of this that the Schur-Constantinescu parameters
have an inheritance property, namely that the parameters of any leading
principal submatrix (recall these will be positive themselves) are the
same as that obtained from the original matrix.

\item {\bf C5} Since the proof of Theorem (\ref{csp}) supplies the Cholesky
factorization of $S$, we get an algorithmic recipe for finding one
Kraus operator representation of a quantum channel $\Phi$. Since the
Cholesky factor, $V$ is lower triangular, the Kraus operators, $V_{i}$,
thereby obtained from $V$ (as described in Section 2), tend to be sparse.
This can be useful in determining sufficient conditions for a channel
to be entanglement breaking, or for computing quantities associated to
channels such as the entanglement fidelity, for instance.
The utility of using the Cholesky factorization lies not just in the
avoidance of spectral calculations (as would be the case if $T$ was found
from the spectral factorization of $S$), but that most of the Kraus operators
$V_{i}$ are then sparse. 
\item {\bf C6} Returning to a positive matrix, $S$, whose entries are scalar,
it is known that if $S_{ii} = 0$, for some $i$, then the entire
row and column to which $S_{ii}$ belongs has to be zero. Therefore,
a reasonable convention to assume is that $\Gamma_{kj} = 0$, whenever
$S_{jj}S_{kk} = 0$. With this convention, the $\Gamma_{kj}, L_{ii}$ provide
a one-one parametrization of positive matrices.  

\item {\bf C7} In the previous section we saw that even the problem
of characterizing pure states via the proposed parametrizations of that
section was not fully resolved. However, the Schur-Constantinescu
{\it parametrization provides a very simple and effective characterization} 
of rank one states, viz., $S$ is rank one iff all $\Gamma_{kj} = 0$, 
except for those
cases in which $S_{jj}S_{kk}\neq 0$, in which case $\Gamma_{kj}$ should
be on the unit circle.

\item {\bf C8} Let $S$ be positive matrix. Then there is a very simple
formula for its determinant in terms of the $\Gamma_{kj}$, viz.,
$$\det (S) =\left(\prod _{k=1}^{d} S _{k,k}\right)\prod _{k<j}
(1-|\Gamma_{kj}|^2).$$ This is useful since some entropic quantities
can often be expressed in terms of determinants, \cite{OP}.

\item {\bf C9} While Equation (\ref{relation}) is intricate, there is
a useful diagram (called a transmission line diagram) which keeps track
of all the matrix products in it. 
\end{itemize}

\section{ Two Further Applications}
In this section two additional applications of the parametrization
of the previous section are provided. The first is to show that
block Toeplitz states have positive partial transpose. The second
is to examine the restrictions on the relaxation rates for an 
open quantum $N$-level system imposed by the requirements of 
complete positivity (cf., \cite{schirmer}).
\subsection{Toeplitz States}
The positive partial trace condition of \cite{Pe,hordecki} has been found to
be a very useful operational condition for entanglement. While, for
general states, it is known to be necessary and sufficient only
for $2\times 2$ and $2\times 3$ states, there have been several arguments
in favour of the notion that states which satisfy this positive
partial trace condition (PPT states) are ``close" to being unentangled, at
least inasmuch as they are not useful for tasks such as dense coding.
Similarly there have been several attempts at studying the PPT property
for positive matrices which satisfy additional conditions, see\cite{BGS}.
In this section we provide a contribution along the same vein. We show
that positive Toeplitz matrices are PPT states. 

The proof of this result was first found by considering the 
Schur-Constantinescu parameters for $3\times 3$  block Topelitz
psoitve matrices. This proof
can be extended in a simple but tedious manner for $d_{i}\times d_{2}$ states.
But there is, in fact, a second proof which works for all dimensions.
We provide this first and then discuss the parameter based proof.

\begin{proposition}
A Toeplitz mixed state is PPT.
\end{proposition}
                                                                                
\noindent
Let $A \in C^{N \times N}$ be a Toeplitz matrix given by
                                                                                
\[
\left[\begin{array}{cccc}
a_0    & a_{-1}  & \cdots & a_{-n}   \\
a_1    & a_{0}   & \cdots & a_{-n+1} \\
\vdots & \vdots  & \ddots & \vdots   \\
a_n    & a_{n-1} & \cdots & a_0
\end{array}\right].
\]
         
We will first, for illustration purposes, show that $A^{T}$ is also
positive. This is, of course, true for arbitrary positive matrices,
but it will serve to illustrate the proof in the partial transpose case. 
\noindent 
Then the $ij$-th entry of $A$ is given by $A_{ij} = a_{i-j}$. 
The transpose of $A$, denoted by $A^T$, is
                                                                                
\[
\left[\begin{array}{cccc}
a_0       & a_{1}  & \cdots & a_{n}   \\
a_{-1}    & a_{0}  & \cdots & a_{n-1} \\
\vdots    & \vdots  & \ddots & \vdots   \\
a_{-n}    & a_{-n+1}  & \cdots & a_0
\end{array}\right].
\]
                                                                                
\noindent ,with $A^T _{ij} = a_{j-i}$. 
Next, in the cycle notation, 
let $\sigma_0$ be the element of the symmetric group $S_N$
on $N$ letters, $\{1, 2, \cdots, N\}$, defined by
                                                                                
\[
\sigma_0 = \prod _{1 \leq k \leq N} (k \; (N-k)).
\]
                                                                                
\noindent $\sigma_0$ induces two simple 
operations on $N \times N$ matrices. 
If $M \in C^{N \times N}$ takes the 
form $M = 
\left[ \begin{array}{c} w_1 \\ w_2 \\ \vdots \\w_n \end{array} \right]$, 
where $w_k$'s are rows of M, we define the operation
$R_{\sigma_0}$ by
                                                                                
\[
M \stackrel{R_{\sigma_0}}{\longrightarrow}
\left[ \begin{array}{c} w_{\sigma_0 (1)} \\ w_{\sigma_0 (2)} \\ \vdots \\w_{\sigma_0(n)} \end{array} \right]
\]
                                                                                
\noindent, i.e. $R_{\sigma_0}$ simply 
permutes the rows of $M$ as specified 
by $\sigma_0$. 
Another operation on columns, $C_{\sigma_0}$, 
is define in the same way. 
Now we notice that if $A$ is Toeplitz as given above, then
                                                                                
\[
\left[ R_{\sigma_0}(C_{\sigma_0}(A)) \right]_{i,j} = A_{N-i, N-j} = a_{j-i} = A^T _{i,j}.
\]
\noindent Since 
$R_{\sigma_0}$ and $C_{\sigma_0}$ preserve the characteristic polynomial, 
we have shown that a if a Hermitian Toeplitz matrix
is positive then so is its transpose.\\

The above fact can be 
extended to the partial transpose 
of an $NM \times NM$ Toeplitz matrix $A$ in the following way: Let
$\sigma_m$ be the same 
permutation as $\sigma_0$ on the 
letters $\{mn, mn+1, \: \cdots \: ,(m+1)n -1 \}$. If $\sigma \in S_{N^2}$
is defined to be the 
disjoint product $\sigma_0 \sigma_1 \cdots \sigma_{M-1}$, and 
$R_{\sigma}$ and $C_{\sigma}$ are the
induced operators, 
then by the same argument as above, 
we have $R_{\sigma}(C_{\sigma}(A)) = A^{PT}$, where
$A^{PT}$ denotes the partial transpose of $A$. 
Once again these operations preserve the characteristic polynomial 
for Toeplitz matrices and hence if  $A$ is positive, in addition, we  
find that so is $A^{PT}$.  
Thus a positive Toeplitz matrix is PPT.\\

The Schur parametrization of 
positive matrices gives another proof of proposition 1 that is immediate.
If $B$ is a  block Toeplitz matrix, 
then $B$ is also Toeplitz. So let $B$
be, for instance, a $3 \times 3$ block Toeplitz matrix. 
Using the Schur-Constantinescu parameters and the block
Toeplitz property of $B$, we can write $B$ explicitly as                                                                                 
\[
\left[\begin{array}{ccc} A    &  A^{\frac{1}{2}} \Gamma_1 A^{\frac{1}{2}} & A^{\frac{1}{2}} (\Gamma_1^2 + D_{\Gamma_1 ^*} \Gamma_2 D_{\Gamma_1}) A^{\frac{1}{2}}  \\ A^{\frac{1}{2}} \Gamma_1 ^* A^{\frac{1}{2}} &  A   &  A^{\frac{1}{2}} \Gamma_1 A^{\frac{1}{2}}\\ A^{\frac{1}{2}}((\Gamma_1^*)^2 + D_{\Gamma_1} \Gamma_2 ^* D_{\Gamma_1 ^*}) A^{\frac{1}{2}}   & A^{\frac{1}{2}} \Gamma_1 ^* A^{\frac{1}{2}} & A \end{array}\right]
\]                                                                                 
\noindent  ,where each entry 
is an $N \times N$ matrix. Note that due to the block-Toeplitz nature
of $B$ its Schur-Constantinescu parameters $\Gamma_{ij}$ need be indexed
by only one subscript.  
Transpose block-wise gives us $A^{PT}$. By the spectral theorem,
$D_{\Gamma_1 ^T} = (D_{{\Gamma_1}}^*) ^T$. So, simply by inspection, we see that $A^{PT}$ has Schur parameters
$\{ (A^{\frac{1}{2}}) ^T , \Gamma_1 ^T, \Gamma_2 ^T \}$. Therefore $A^{PT} \geq 0$. This is in fact true in general:
                                                                                
\begin{proposition}
If $A \in C^{MN \times MN}$ is block Toeplitz, then $A$ is PPT.
\end{proposition}

The basic idea is to show that
If $A$ is parametrized by $\{\Gamma_i \}$, 
then $A^{PT}$ is parametrized by $\{ \Gamma _i ^T\}$. Note that the
block-Toeplitz property means that Schur-Constantinescu parameters
of $A$ depend only on one index (cf., the $3\times 3$ block case). 
We will omit the proof, which is straightforward but tedious. 
Via the combinatorial structure of the Schur parameters,
one can see how the parametrization of $A$ gives rise to that of $A^{PT}$.
The so-called "lattice structure" of the Schur parameters 
for the $4 \times 4$ case is shown in Figure 1 below.
Each transfer box
in Figure 1 describes the action of the Julia operator  
$U(\Gamma _i)$.

\noindent Let $U^T(\Gamma)$ denote 
the transpose of the Julia operator of $\Gamma$, i.e.                                                                                 
\[
U^T (\Gamma) =
\left[\begin{array}{cc}
\Gamma^T            &  (D_{\Gamma})^T  \\
(D_{\Gamma ^*})^T   & - (\Gamma ^*)^T
\end{array}\right]
=
\left[\begin{array}{cc}
\Gamma^T            &  D_{{\Gamma}^{T*}}  \\
(D_{\Gamma ^T})     &  -(\Gamma ^T)^*
\end{array}\right]
=
U(\Gamma ^T).
\]

\noindent 
Each entry of the positve 
semidefinite kernel $\{ A_{ij} \}$ corresponds to 
those paths in the diagram that startfrom $L_{jj}$
and end at ${L_{ii}}^*$. 
For example, each path from $L_{33}$ to ${L_{11}}^*$ 
describes to a summand in the expression for $A_{13}$.
So we can see that the transmission line diagram 
of $A^{PT}$ is then obtained by replacing each $U(\Gamma_i)$ 
transfer box by that of $U^T(\Gamma_i)$.

\begin{figure}[h]
\setlength{\unitlength}{3000sp}%
\begingroup\makeatletter\ifx\SetFigFont\undefined%
\gdef\SetFigFont#1#2#3#4#5{%
  \reset@font\fontsize{#1}{#2pt}%
  \fontfamily{#3}\fontseries{#4}\fontshape{#5}%
  \selectfont}%
\fi\endgroup%
\begin{picture}(6324,2874)(289,-2323)
{ \thinlines
\put(601,-361){\circle{300}}
}%
{ \put(2101,-361){\circle{300}}
}%
{ \put(3901,-361){\circle{300}}
}%
{ \put(5701,-361){\circle{300}}
}%
{ \put(301,-361){\line( 1, 0){900}}
}%
{ \put(601,-211){\line( 0,-1){300}}
}%
{ \put(1201,-361){\vector( 1,-1){600}}
}%
{ \put(1201,-961){\vector( 1, 1){600}}
}%
{ \put(451,239){\framebox(300,300){\small  $L_{44}$}}
}%
{ \put(601,239){\vector( 0,-1){450}}
}%
{ \put(1051,-361){\line( 0, 1){300}}
}%
{ \put(1051,239){\vector( 0, 1){300}}
}%
{ \put(901,-61){\framebox(300,300){\small $L^*_{44}$}}
}%
{ \put(1126,-1111){\framebox(750,900){}}
}%
{ \put(901,-961){\line( 1, 0){1200}}
}%
{ \put(1201,-361){\line( 1, 0){1200}}
}%
{ \put(2926,-1111){\framebox(750,900){}}
}%
{ \put(2026,-1711){\framebox(750,900){}}
}%
{ \put(3826,-1711){\framebox(750,900){}}
}%
{ \put(4726,-1111){\framebox(750,900){}}
}%
{ \put(2926,-2311){\framebox(750,900){}}
}%
{ \put(2101,-961){\line( 1, 0){1200}}
}%
{ \put(3301,-961){\line( 1, 0){1200}}
}%
{ \put(4501,-961){\line( 1, 0){1200}}
}%
{ \put(2401,-361){\line( 1, 0){1200}}
}%
{ \put(3601,-361){\line( 1, 0){1200}}
}%
{ \put(4801,-361){\line( 1, 0){1200}}
}%
{ \put(1201,-361){\vector( 1, 0){600}}
}%
{ \put(1201,-961){\vector( 1, 0){600}}
}%
{ \put(3001,-361){\vector( 1, 0){600}}
}%
{ \put(3001,-961){\vector( 1, 0){600}}
}%
{ \put(2101,-961){\vector( 1, 0){600}}
}%
{ \put(2101,-1561){\vector( 1, 0){600}}
}%
{ \put(3001,-1561){\vector( 1, 0){600}}
}%
{ \put(3001,-2161){\vector( 1, 0){600}}
}%
{ \put(3901,-961){\vector( 1, 0){600}}
}%
{ \put(3901,-1561){\vector( 1, 0){600}}
}%
{ \put(4801,-361){\vector( 1, 0){600}}
}%
{ \put(4801,-961){\vector( 1, 0){600}}
}%
{ \put(3001,-961){\vector( 1, 1){600}}
}%
{ \put(4801,-961){\vector( 1, 1){600}}
}%
{ \put(3001,-361){\vector( 1,-1){600}}
}%
{ \put(4801,-361){\vector( 1,-1){600}}
}%
{ \put(2101,-1561){\vector( 1, 1){600}}
}%
{ \put(3901,-1561){\vector( 1, 1){600}}
}%
{ \put(3001,-2161){\vector( 1, 1){600}}
}%
{ \put(2176,-961){\vector( 1,-1){600}}
}%
{ \put(3001,-1561){\vector( 1,-1){600}}
}%
{ \put(3901,-961){\vector( 1,-1){600}}
}%
{ \put(1801,-1561){\line( 1, 0){3000}}
}%
{ \put(2701,-2161){\line( 1, 0){1200}}
}%
{ \put(2101,-511){\line( 0, 1){300}}
}%
{ \put(3901,-511){\line( 0, 1){300}}
}%
{ \put(5701,-511){\line( 0, 1){300}}
}%
{ \put(1951,239){\framebox(300,300){\small $L_{33}$}}
}%
{ \put(3751,239){\framebox(300,300){\small $L_{22}$}}
}%
{ \put(5551,239){\framebox(300,300){\small $L_{11}$}}
}%
{ \put(2101,239){\vector( 0,-1){450}}
}%
{ \put(3901,239){\vector( 0,-1){450}}
}%
{ \put(5701,239){\vector( 0,-1){450}}
}%
{ \put(2401,-61){\framebox(300,300){\small $L^*_{33}$}}
}%
{ \put(4201,-61){\framebox(300,300){\small $L^*_{22}$}}
}%
{ \put(2551,-361){\line( 0, 1){300}}
}%
{ \put(4351,-361){\line( 0, 1){300}}
}%
{ \put(2551,239){\vector( 0, 1){300}}
}%
{ \put(4351,239){\vector( 0, 1){300}}
}%
{ \put(6001,-511){\framebox(300,300){\small $L^*_{11}$}}
}%
{ \put(6376,-361){\vector( 1, 0){225}}
}%
\end{picture}

\caption{ Lattice structure for $4\times 4$ positive matrices}
\end{figure}
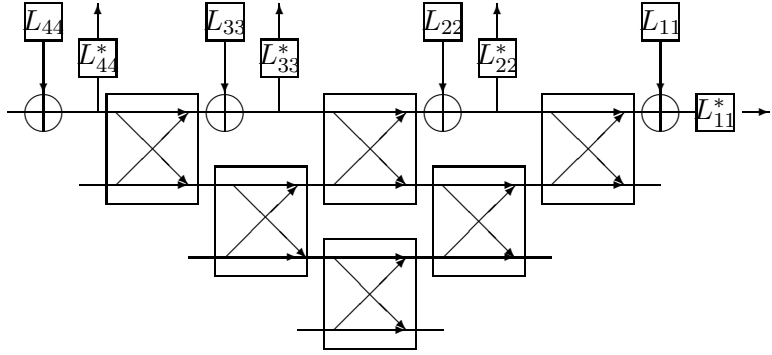

\subsection{Constraints on Relaxation Rates}
In this subsection we revisit the very interesting work of
\cite{schirmer} on the constraints imposed on the relaxation
rates of an open $N$-level quantum system by the requirement
that its evolution be completely positive. {\it In order to
keep the notation the same as in \cite{schirmer}, we will,
in this subsection only denote the Schur-Constantinescu parameters
by $g_{ij}$ (and not $\Gamma_{ij}$).} 
 
Let us first briefly review the contents of \cite{schirmer}.
Let $\rho (t)$ be the state of an open $N$-level quantum
system and let $\tilde{\rho}$ be the vector in $C^{N^{2}}$
which represents ${\mbox vec} (\rho )$. Then its evolution
can be expressed via
\begin{equation}
\label{lindblad} 
\dot{\tilde{\rho }} = (-\frac{i}{\hbar}L_{H} + L_{D})\tilde{\rho }
\end{equation}
 
where $L_{H}$ and $L_{D}$ are $N^{2}\times N^{2}$ matrices representing
the Hamiltonian and dissipative parts respectively of the evolution
of $\tilde{\rho}$. Let the index $(m,n)$ denote the number $m + (n-1)N$.
Then the non-zero entries of $L_{D}$ are given by
\begin{eqnarray*}
(L_{D})_{(m,n)(m,n)} & = & -\Gamma_{mn}, m\neq n\\
(L_{D})_{(m,m)(l,l)} & = & \gamma_{ml}, m\neq l\\
(L_{D})_{(m,m)(m,m)} & = & -\sum_{k=1,k\neq m}^{N}\gamma_{km}
\end{eqnarray*}  
 Here $\gamma_{kn}$ is the population relaxation rate from level
$\mid n >$ to $\mid k >$. The $\gamma_{kn}$ are real and non-negative.
$\Gamma_{kn}$ (for $k\neq n$) is the dephasing rate for the
transition from $\mid k >$ to $\mid n >$. Since, $\Gamma_{kn} = \Gamma_{nk}$,
it is easily seen that $(L_{D})_{(m,n)(m,n)} =
(L_{D})_{(n,m)(n,m)}$. A key step in the work of \cite{schirmer} is to
express $\Gamma_{kn}$ as a sum of two summands, in recognition of the
fact that dephasing is also enhanced by population relaxation, to wit
\[
\Gamma_{kn} = \Gamma_{kn}^{p} + \Gamma_{kn}^{d}
\]
with $\Gamma_{kn}^{p}$, the decoherence rate due to population relaxation
and  $\Gamma_{kn}^{d}$ the decoherence rate due to pure phase relaxation.   
The requirement that the open quantum system's evolution be completely
positive, \cite{alicki,sudarshan}, 
imposes restrictions on $\gamma_{kn}$ and $\Gamma_{kn}$. These restrictions
can be expressed as the requirement that a certain $(N^{2}-1)\times
(N^{2} - 1)$ matrix concocted out of the $\gamma_{kn}$ and $\Gamma_{kn}$
be positive, \cite{schirmer}. However, per \cite{schirmer}, this requirement
can be reduced to verifying that a related $(N-1)\times (N-1)$ matrix
be positive. The form of this $(N-1)\times (N-1)$ matrix will depend
on a choice of an orthogonal basis for the space of traceless, Hermitian 
$(N^{2}-1)\times
(N^{2} - 1)$ matrices. However, positivity of this matrix itself is
independent of the choice of basis. The excellent analysis of \cite{schirmer}
is unfortunately
marred for the $N = 4$ case by an incorrect criterion for positivity.
Indeed, Equation (28) of \cite{schirmer} are only necessary for
positivity, while Equations (31)-(32) are (as correctly claimed in
\cite{schirmer}) also just necessary (though they come closer to
sufficiency than Equation (28) of \cite{schirmer}). 

In the sequel, we will use the Schur-Constantinescu parameters to analyse
the $N=4$ case of \cite{schirmer}. As in \cite{schirmer} the evolution
Equation (\ref{lindblad}) is completely positive iff the $3\times 3$ real
symmetric matrix $B = (b_{ij})$ is positive. To specify the entries
of the $b_{ij}$, we denote by $\Gamma^{d}_{tot}$ the quantity
$\frac{1}{2}\sum_{n=2}^{4}\sum_{m=1}^{n-1}\Gamma^{d}_{mn}$. Then
the entries of $B$ are given by
\begin{eqnarray*} 
b_{11} & = & \Gamma^{d}_{tot} - (\Gamma_{13}^{d} + \Gamma_{24}^{d}\\ 
b_{22} & = & \Gamma^{d}_{tot} - (\Gamma_{13}^{d} + \Gamma_{24}^{d}\\
b_{33} & = & \Gamma^{d}_{tot} - (\Gamma_{12}^{d} + \Gamma_{34}^{d}\\   
b_{12} & = & \frac{(\Gamma_{12}^{d} - \Gamma_{34}^{d})}{2}\\
b_{13} & = & \frac{(\Gamma_{14}^{d} - \Gamma_{23}^{d})}{2}\\  
b_{23} & = & \frac{(\Gamma_{13}^{d} - \Gamma_{24}^{d})}{2}
\end{eqnarray*}

Now $B$ is positive iff $b_{ii}\geq 0, i=1,\ldots , 3$ and
the Schur-Constantinescu parameters $g_{12}, g_{13}, g_{23}$ are in
the closed unit disc. Since $B$ is real this is equivalent to demanding
that the $g_{ij}$ belong to the interval $[-1, 1]$.

The conditions $b_{ii}\geq 0$ become
\begin{eqnarray*}
\Gamma_{12}^{d} + \Gamma_{14}^{d} + \Gamma_{23}^{d} + \Gamma_{34}^{d}
\geq \Gamma_{13}^{d} + \Gamma_{24}^{d}\\
\Gamma_{12}^{d} + \Gamma_{13}^{d} + \Gamma_{24}^{d} + \Gamma_{34}^{d}
\geq \Gamma_{14}^{d} + \Gamma_{23}^{d}\\   
\Gamma_{13}^{d} + \Gamma_{14}^{d} + \Gamma_{23}^{d} + \Gamma_{24}^{d}
\geq \Gamma_{12}^{d} + \Gamma_{34}^{d}
\end{eqnarray*}

Now $b_{12} = \sqrt{b_{11}}g_{12}\sqrt{b_{22}}$. So $g_{12}\in [-1,1]$
becomes 
\[
4\Gamma_{12}^{d}\Gamma^{d}_{34} - (\Gamma_{13}^{d} - \Gamma_{14}^{d})^{2}
- (\Gamma_{13}^{d} - \Gamma_{23}^{d})^{2} +
(\Gamma_{13}^{d} - \Gamma_{24}^{d})^{2} + 
(\Gamma_{14}^{d} - \Gamma_{23}^{d})^{2} -
(\Gamma_{14}^{d} - \Gamma_{24}^{d})^{2} -
(\Gamma_{23}^{d} - \Gamma_{24}^{d})^{2} \geq 0
\]

Likewise the condition $g_{23}\in [-1,1]$ becomes
\[
4\Gamma_{13}^{d}\Gamma^{d}_{24} - (\Gamma_{12}^{d} - \Gamma_{14}^{d})^{2}
- (\Gamma_{12}^{d} - \Gamma_{23}^{d})^{2} + 
(\Gamma_{12}^{d} - \Gamma_{34}^{d})^{2} +
(\Gamma_{14}^{d} - \Gamma_{23}^{d})^{2} -
(\Gamma_{14}^{d} - \Gamma_{34}^{d})^{2} -
(\Gamma_{23}^{d} - \Gamma_{34}^{d})^{2} \geq 0
\]
   
Finally $g_{13} \in [-1,1]$ becomes
\[
b_{11}b_{22}b_{33} + 2b_{12}b_{13}b_{23}\geq b_{11}b_{23}^{2}
+ b_{22}b_{13}^{2} + b_{33}b_{12}^{2}
\]
Note that the condition $\mid g_{13}\mid \leq 1$ is not similar
to the condition for the other $g_{ij}$ to be in $[-1,1]$. This is
to be expected since the formula for $g_{jk}$ for $k > j+1$ is more
intricate than those for the $g_{jk}, k = j+1$.
Furthermore, this last condition is precisely one of those obtained in 
\cite{schirmer}. However, the conditions obtained here are necessary
and sufficient.

\section{Conclusions}
Since positive matrices play a vital role in many applications, it is
of importance to obtain computable parametrizations of them. In this
paper we discussed several such potential parametrizations. Which
one of them one ought to use is, of course, a matter dictated by
the application one has in mind. We argued, hopefully persuasively,
in favour of the versatility of the parametrization proposed in \cite{tivin}.
There are several other applications besides the ones discussed here,
to which one could apply this parametrization. This will be the
subject of future work.

\end{document}